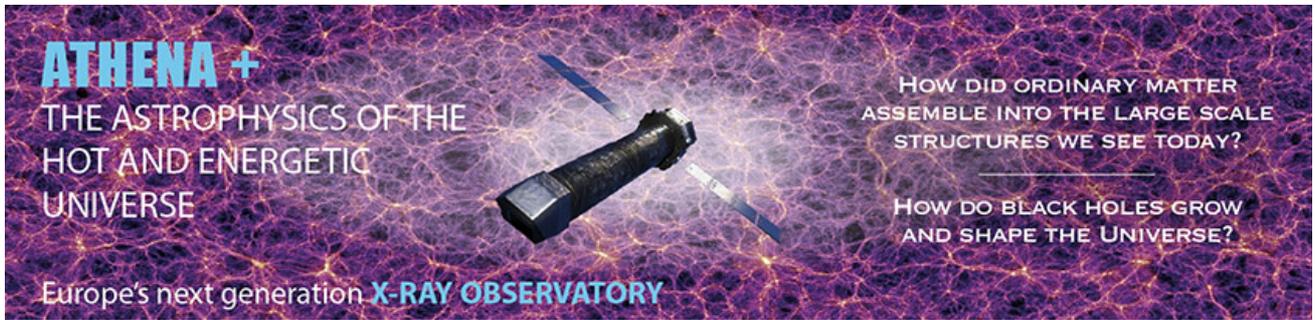

# The Hot and Energetic Universe

An *Athena+* supporting paper

# AGN Feedback in Galaxy Clusters and Groups

## Authors and contributors

**J.H. Croston, J.S. Sanders**, S. Heinz, M.J. Hardcastle, I. Zhuravleva, L. Bîrzan, R.G. Bower, M. Brüggen, E. Churazov, A.C. Edge, S. Ettori, A.C. Fabian, A. Finoguenov, J. Kaastra, M. Gaspari, M. Gitti, P.E.J. Nulsen, B.R. McNamara, E. Pointecouteau, T.J. Ponman, G.W. Pratt, D.A. Rafferty, T.H. Reiprich, D. Sijacki, D.M. Worrall, R.P. Kraft, I. McCarthy, M. Wise

# 1. EXECUTIVE SUMMARY

Active galactic nuclei (AGN) in the centres of galaxy groups and clusters are believed to play a crucial role in galaxy formation and cluster evolution, by regulating the cooling of the hot intragroup and intracluster gas. Over the past decade it has become increasingly clear that mechanical feedback via AGN jets is one of the best candidates for the suppression of star formation in the most massive galaxies, potentially reconciling the predictions of galaxy formation models with the properties of the observed galaxy population. The current generation of X-ray observatories, *Chandra* and *XMM-Newton*, have revealed a complex interplay between cooling and heating, via spectroscopic measurements of gas cooling and detailed imaging of the interactions of jets and gas in cluster cores. However, the physics of how the balance between these processes is established and maintained can only be glimpsed in a few nearest systems and so remains poorly understood. At the same time, the role of jet heating at higher redshifts, and its relative significance compared to quasar (radiative) feedback as a function of epoch and environment, remains unclear. Key outstanding questions include:

- How is the energy from jets dissipated and distributed throughout the hot gas atmosphere of the cluster or group?
- How does feedback operate to regulate gas cooling and AGN fuelling?
- What is the cumulative impact of powerful radio galaxies on the evolution of baryons from the epoch of group and cluster formation to the present time?

**To answer these questions, a major advance in X-ray astronomy capability is required. Only the large collecting area, high spectral and spatial resolution of a mission such as *Athena+* can provide the breakthrough diagnostic ability necessary for developing a robust physical understanding of AGN jet feedback. This will be achieved via exquisitely detailed studies of low-redshift groups and clusters, and via the removal of biases through dramatically improved statistics for population studies of jet feedback in groups and clusters to $z\sim1$.**

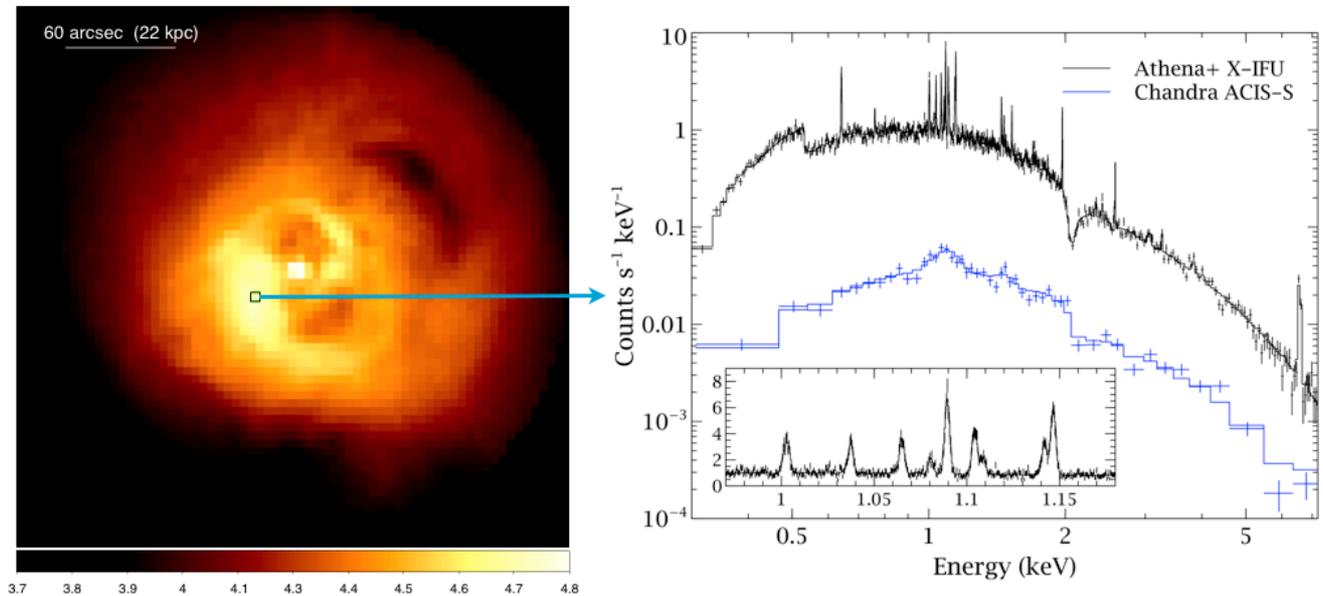

Fig. 1. Simulated 50-ks X-IFU observation of the centre of the Perseus cluster (0.5 to 7 keV). The left panel shows the log number of counts per square arcsecond. The spectrum on the right is from the single 5"×5" region marked by the box, with a Chandra spectrum of the same observation length for comparison. The inset shows the region around the iron L complex. The velocity broadening can be measured to 10-20 km s$^{-1}$, the temperature to 1.5% and the metallicity to 3% on scales <10kpc in 20-30 nearby systems, and on <50kpc scales in hundreds of clusters and groups.

## 2. HOW IS JET ENERGY DISSIPATED AND DISTRIBUTED?

The gross energetics of AGN feedback in local hot atmospheres (groups and clusters) are reasonably well established. Remarkably, weak radio sources at the centres of nearby clusters often have mechanical power comparable to the radiative output of a quasar, which is sufficient to prevent the majority of material in the hot atmospheres from cooling (e.g. McNamara & Nulsen 2007). The heat source — the accreting black hole — is roughly the size of the Solar System, yet the rate of mechanical energy input must be tuned to conditions operating over scales nine or more decades larger. Similar feedback processes must act to suppress cooling in systems ranging from massive elliptical galaxies to galaxy groups and clusters, and observations of X-ray cavities in all types of system provide observational support for this picture. It is only in the nearest bright cluster cores, however, that the first clues have been found as to how the jet power, originally highly collimated, may be isotropically distributed to the surrounding gas: quasi-spherical ripples are observed in the X-ray emission that are interpreted as sound waves and weak shocks (e.g. Fabian 2012). In this interpretation, enough energy to offset cooling is carried by these disturbances, but the microphysics of how much energy is dissipated in the gas, and over which spatial scales, is not understood.

Understanding the dynamics and heating mechanisms in these powerful outflows demands a leap in spatially-resolved spectral resolution by more than an order of magnitude above *Chandra* and *XMM-Newton*. *Athena+* will enable the first detailed mapping of the velocity field of the hot gas to an accuracy of tens of km s$^{-1}$ (Figs. 1, 2 and 3), allowing the viscosity and dissipation mechanisms to be determined. In these nearest systems it will be possible to do complete calorimetry of energy input by AGN for the first time, adding up the energy stored in cavities, ripples, and motions, and following its evolution as it is transferred from the jets to the environment in the inner core and then escapes out to larger radii. From accurate measurements of line profiles and from variations of the line centroid over the image it is possible to deduce the characteristic spatial scales and velocity amplitude of large (> kpc) turbulent eddies (e.g., Zhuravleva et al. 2012), while the total width of the line provides a measure of the total kinetic energy stored in the stochastic gas motions at all spatial scales. Such measurements of the kinematics of the hot gas, which absorbs the bulk of the jet energy, and contains the bulk of the gaseous mass, are only possible at X-ray wavelengths and with the spectral and spatial resolution of *Athena+*.

*Athena+* will not only be a unique instrument to establish the microphysics of feedback in the nearest systems, but will also enable the first population studies of AGN-induced perturbations over a broad range of systems and spatial scales. With *Athena+*'s collecting area and the high spatial resolution of the WFI instrument it will be possible to detect and characterize ripples and disturbances in the X-ray surface brightness distribution of 40-50 nearby clusters and groups (Fig. 3). It will therefore be possible to establish whether such features are ubiquitous, and to relate the mechanical energy that they transport to environmental and AGN properties for the first time across a wide mass range. In addition, *Athena+* will be much more sensitive to cavities in the X-ray emission further from the cluster core, allowing us to measure the energetics of multiple cycles of feedback in a consistent way. A few giant cavity systems observed with *Chandra* (e.g. MS0735+7421, Hydra A) provide an intriguing glimpse of the potential importance of AGN heating of cluster gas on scales beyond the cluster core, affecting entropy distributions and scaling relations (e.g. Gitti et al. 2007, and see the final section).

*Athena+* will directly measure or place restrictive limits on the thermal and non-thermal content of X-ray cavities (e.g. Sanders & Fabian 2007, Croston et al. 2008), helping to establish their initial composition and the subsequent role of entrainment of the intracluster gas. Together with a greatly improved ability to study strong shocks (see final section), *Athena+* will establish the relative importance of different heating mechanisms and how they vary across the population. Crucially for the interpretation of next-generation radio surveys, this will greatly improve the calibration of radio luminosity – jet power relations (e.g. Bîrzan et al. 2008), which do not currently take into account the energy stored in sound waves and shocks. Further improvements in these relations will be obtained from large sample studies including X-ray cavity detections and velocity measurements out to $z\sim 1$.

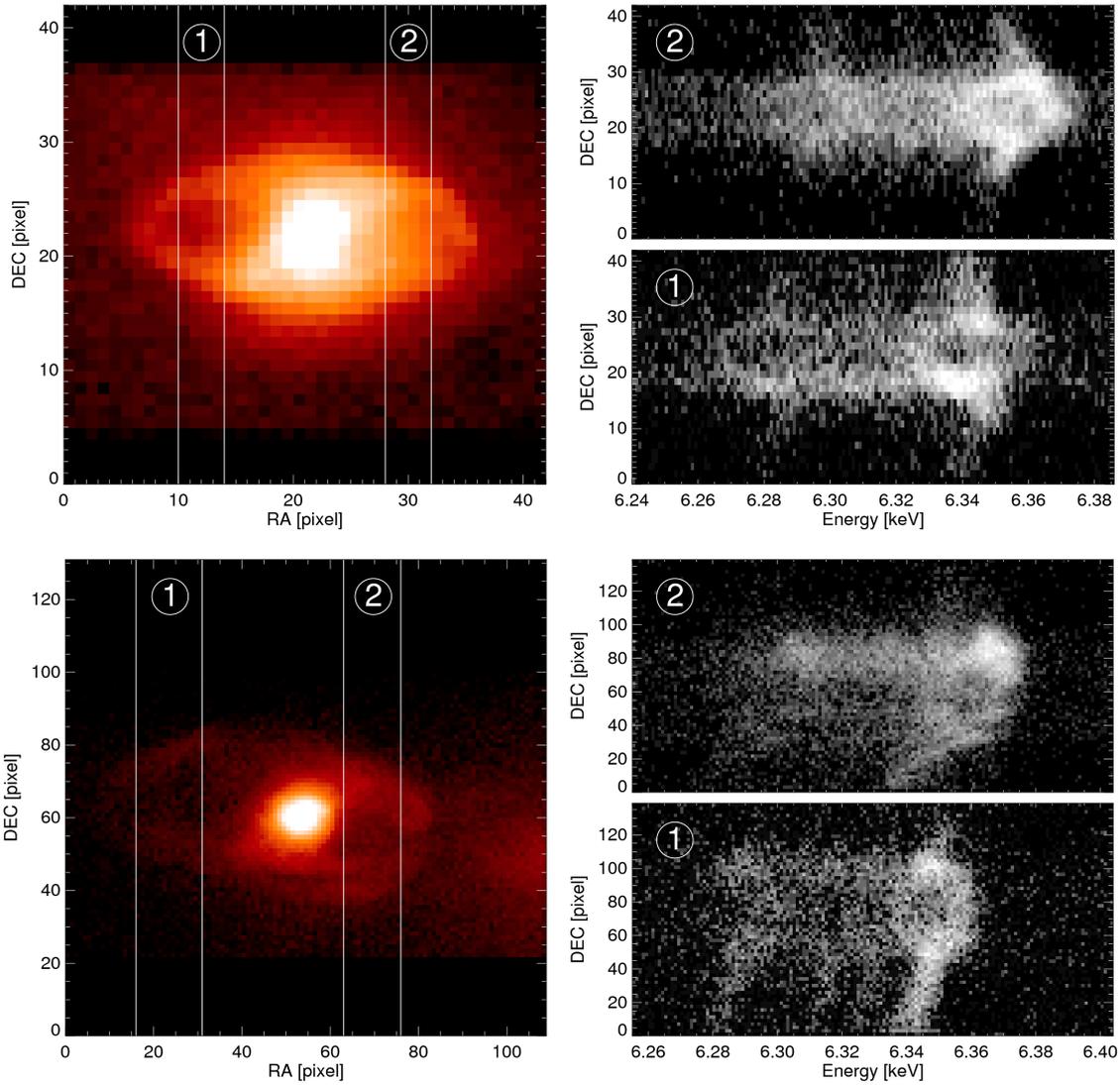

Fig. 2. Simulated high-resolution X-ray spectra from the shells and X-ray cavities of Cygnus A (top) and Hydra A (bottom), with ages of 21 and 170 Myr, respectively, based on hydrodynamical simulations of Heinz et al. (2010). X-ray images and chosen virtual spectral slits (left); right: spatially resolved spectra of the 6.7 keV Fe XXV K$\alpha$ line as observed by the *Athena+* X-IFU in 250 ks. At the location of the cavities, each line splits into approaching, restframe and receding components from which the velocity, age and therefore the jet power can be derived.

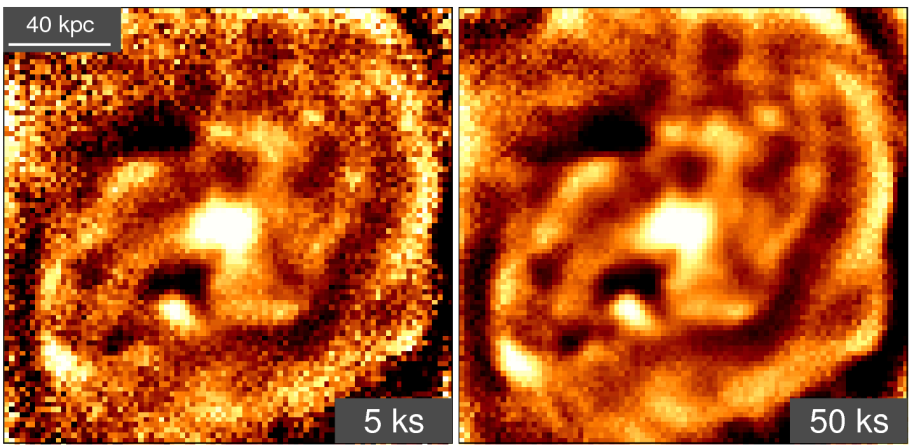

Fig. 3. Unsharp mask images made from 5 and 50ks exposure simulated WFI observations of a cluster core at $z$=0.05 based on the simulations of Morsony et al. (2010), demonstrating the ability to detect and characterize ripples and weak shocks in 40-50 groups and clusters over a wide mass range, allowing the mechanical energy they transport to be related to environmental and AGN properties for the first time.

A further important effect of radio lobe evolution on the hot gas is the uplift and mixing of material to large radii by AGN jets and rising buoyant bubbles. X-ray imaging from *XMM-Newton* and *Chandra* has shown cool, metal enriched plumes of gas lying along the radio sources and X-ray bubbles of many clusters, with heavy elements distributed anisotropically and aligned with the large-scale radio and cavity axes (e.g. Rasmussen et al. 2009, Kirkpatrick et al. 2011), as is also suggested by simulations (e.g. Gaspari et al. 2011). The amount of gas being displaced lies upward of $10^{10}$ solar masses, with implied outflow rates of tens to hundreds of solar masses per year. This rivals the level of cooling expected in the cluster core, and thus may be an important mechanism affecting galaxy formation at late times. The X-IFU will be able to measure and map the outflow speeds, expected to be hundreds of km s$^{-1}$, and flow patterns. A significant fraction of the outflowing gas should fall back to the center of the cluster in a fountain, which may also be mapped with the X-IFU. Detailed metal distribution mapping also provides a powerful tracer of jet energy distribution, AGN-induced turbulence and the history of such gas motions (e.g. Rebusco et al. 2005). *Athena+* studies of the relationship between metallicity and jet activity in nearby clusters will therefore provide important tests of metallicity evolution in cosmological simulations (see Pointecouteau, Reiprich, et al., 2013 and Ettori, Pratt, et al., 2013, *Athena+* supporting papers).

## 3. HOW DOES FEEDBACK OPERATE TO REGULATE GAS COOLING AND AGN FUELLING?

The jets supplying the energy thought to regulate cooling of the hot intracluster and intragroup medium, and thus star formation in massive galaxies, are powered by the gravitational energy released by material falling onto the central AGN. The origin of the AGN fuel and its accretion mechanism are, however, not yet understood. The persistence of short central cooling times in a large fraction of hot gas atmospheres, together with the absence of cold material and star formation at the level this implies, is difficult to explain unless jet heating is self-regulated in a feedback loop (e.g. McNamara & Nulsen 2007). This requires the fuelling process to be closely linked to the thermodynamical properties of the hot gas that absorbs most of the jet's impact. There are (at least) two ways that the hot gas phase could fuel the AGN and power the observed radio jets: Bondi (1952) accretion from the hot atmosphere itself could provide a steady supply of fuel, but another, perhaps more likely, fuel source is cold molecular gas that has cooled out of the hot gas (e.g. Pizzolato & Soker 2005; Sharma et al. 2012; Gaspari et al. 2012). For the same black hole mass and hot gas properties these models predict substantially different accretion rates, and so it is vital that the fuelling process is understood. Progress in testing accretion models and in confirming the ability of jet feedback to regulate cooling and star formation relies on a complete physical understanding of the cooling process across the temperature range, which can be provided for the first time by *Athena+*.

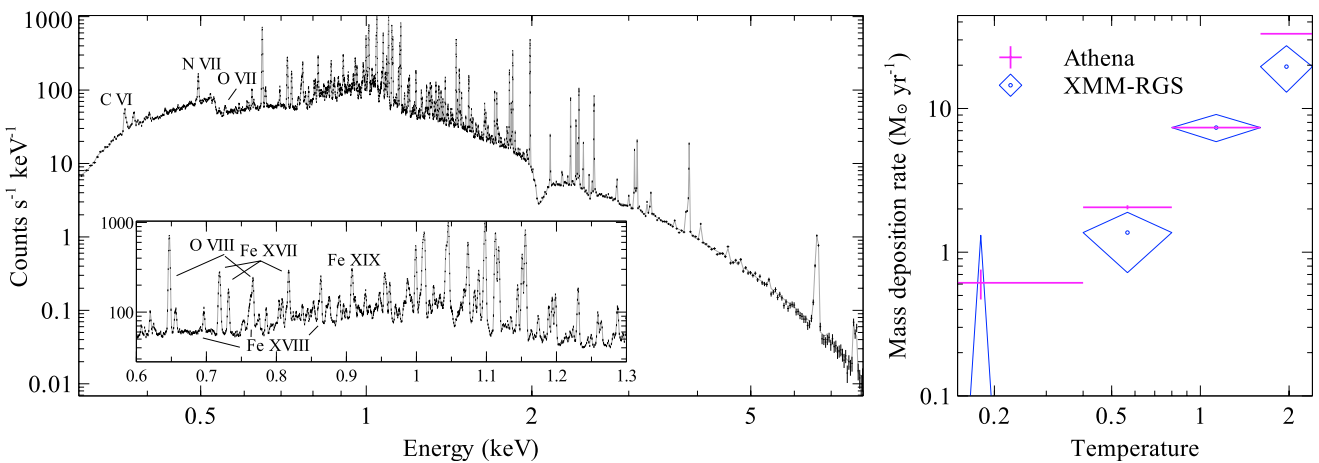

Fig. 4. (Left) 20-ks simulated X-IFU spectrum of the Centaurus cluster core by *Athena+*. A large number of X-ray emission lines can be seen in this short observation (see zoom up of Fe-L region). We label some emission lines sensitive to the coolest X-ray emitting material. (Right) Comparison of constraints on mass deposition rate as a function of gas temperature achievable in 20-ks exposures with *Athena+* X-IFU and *XMM*-RGS exposures, showing more than two orders of magnitude improvement in efficiency.

Despite the undoubtedly important energy input from jets, the clusters with the shortest radiative cooling times show clear evidence for substantial amounts of cool material: optical line-emitting nebulae (e.g. Crawford et al. 1999), often associated with dust, coexist with large masses of atomic and molecular gas ($\sim 10^{11}$ solar masses) at much lower temperatures, as seen in CO (e.g. Edge 2001) and $H_2$ emission (e.g. Jaffe et al. 2001). In the next few years ALMA will enable detailed kinematic studies of the filaments via CO emission, while *Herschel* and in the future *SPICA* and *JWST* will provide further advances in our knowledge of the cold material in central cluster galaxies. Star formation is also observed in many objects (e.g. O'Dea et al. 2008). The filaments are observed to be dragged out by the rising bubbles (Fabian et al. 2003), and are also visible by their soft X-ray emission. These nebulae are clearly an intricate part of the heating and cooling interplay in cluster cores, and a detailed characterization of their thermodynamics and evolution is essential for a complete understanding of how star formation in the most massive galaxies (typically found in the centres of clusters and groups) is regulated.

Galaxy clusters are the best environment to measure material over a wide temperature range, and the X-ray-emitting regime is where the bulk of the energy needs to be lost for material to cool. Although cold gas is present, *XMM*-RGS observations have shown that there is much less cool X-ray emitting material in the centres of these objects than would be expected from radiative cooling alone (e.g. Peterson et al. 2003), requiring AGN feedback. However, deep observations have found smaller amounts of X-ray emitting material down to around 5 million K (e.g. Sanders et al. 2008). The cooling timescale of such gas is only $\sim 10^7$ yr; thus the feedback cycle must operate on short timescales to completely suppress cooling. Such deep observations are only possible in a few cases, and cooling rates cannot currently be measured down to the levels of observed star formation. *Athena+* will make vast improvements to the measurement of the X-ray temperature distribution in the cores of clusters and groups, enabling the first detailed comparisons of X-ray cooling rates with the observed star formation rates. Fig. 4 demonstrates the spectral quality achievable in short exposures, with the ability to measure cooling rates below 4 million K, which would require more than two orders of magnitude deeper *XMM*-RGS exposures. Thus a much larger sample of objects can be examined in great depth, and the time evolution of the heating-cooling balance can be investigated for the first time.

An interesting additional method for measuring the cold gas mass independently of the thermal state of the gas comes from measuring the 6.4 keV fluorescent iron line equivalent width (Churazov et al. 1998), achievable with *Athena+* for the first time. The X-IFU's high spatial resolution will also allow the temperature distribution to be measured on a point-to-point basis and the X-ray emission mechanism of the filaments (a crucial diagnostic of the physics of the multi-phase regions, e.g. Fabian et al. 2011) to be determined. X-IFU measurements of the dynamics of the hot gas in the vicinity of cool filaments can be used to investigate whether their motions are correlated, helping to distinguish locations where filaments are being evaporated by the hot gas, where they are thermally unstable to cooling (potentially important for accretion – see below) and where mixing is occurring. *Athena+* will therefore enable the first detailed accounting of the fraction of material cooling out of the hot-gas phase that is forming stars, and how much is available for AGN fuelling.

As well as providing the first tight constraints on hot gas cooling rate across the group and cluster population, *Athena+* will also map the turbulent velocities in cluster and group cores over a range of jet power, which will confirm the importance of AGN-induced turbulence. This process could play an important role in driving the growth of thermal instabilities leading to hot gas condensation in the very inner regions of the cluster, which is one possible AGN fuelling model (e.g. Gaspari et al. 2012). Finally, the complete calorimetry of the jet energy input that will be possible with *Athena+* (see first section) will lead to robust jet power estimates for very large samples (hundreds of groups and clusters). Hence accretion rates from cold and hot material can then be compared directly with well determined jet power measurements, making it possible to establish which fuelling modes dominate in which environments and whether black holes grow in accordance with the Bondi prescription as typically assumed in cosmological simulations (e.g. Sijacki et al. 2007). Such comparisons are the only way to confirm the existence of a self-regulating feedback process operating to regulate cooling and star formation in the most massive galaxies.

## 4. WHAT IS THE CUMULATIVE IMPACT OF POWERFUL RADIO GALAXIES?

In the cores of nearby galaxy clusters and groups, mechanical feedback appears tightly coupled to cooling, with more or less continuous feedback required to offset the short radiative cooling times. This feedback cannot rely on strong shock heating, as observations demonstrate the persistence of steep abundance gradients that would be disrupted by violent heating processes. However, in the less rich environments typical for the nearest powerful radio galaxies (e.g. Ineson et al. 2013), the jets must be transferring significantly more energy to the environment than is required to offset cooling. This type of AGN heating is likely to have been common at the epoch where groups and clusters formed, and could explain the origin of the entropy excess observed in groups and clusters (e.g. Pratt et al. 2010, Ma et al. 2013, and see Ettori, S., Pratt, G.W., et al., 2013, Athena+ supporting paper); however, it is poorly understood due to long-standing difficulties in determining radio-galaxy physical conditions and dynamics via radio synchrotron emission, and the limitations of observing group-scale environments at moderate redshifts with *Chandra* and *XMM-Newton*.

Numerical models of radio-galaxy evolution can investigate energetic impact, and the role of shock-heating and other mechanisms as a function of jet power, environment and source age (e.g. Heinz et al. 2010, Hardcastle & Krause 2013), but X-ray observations provide the only means of testing these models by directly detecting shocked gas surrounding expanding radio lobes. With *Chandra* and *XMM-Newton* this has only been possible in very few systems with poor constraints on shock speeds, whereas *Athena+* will be able to obtain precise measurements of shock conditions in large samples of objects at intermediate redshifts. Next-generation radio surveys (e.g. LOFAR, SKA) will revolutionize our knowledge of the radio-galaxy population, in particular revealing the population of 'typical' low and moderate luminosity sources to the highest redshifts, but cannot be used to draw robust inferences about AGN jet impact and its cosmological evolution without the crucial tests of radio-galaxy dynamical models that can only be provided by *Athena+*.

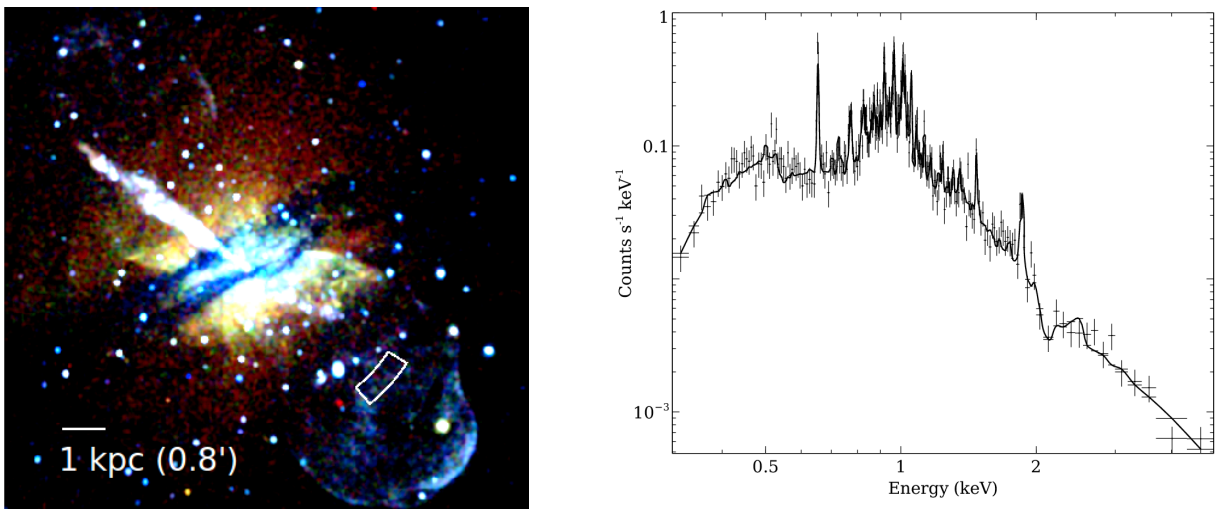

Fig. 5. Simulated WFI pseudo-colour image (l) and X-IFU spectrum (r) from the region indicated, for a 50-ks observation of Centaurus A, demonstrating *Athena+*'s ability to obtain the first direct measurements of advance speed for a strong radio-lobe shock. The shock speed can be determined to within 10% via measurements of line broadening from small regions of the X-ray shell emission dominated by thermal emission (Croston et al. 2009).

X-IFU's spectral resolution will enable direct measurement of shock expansion speeds in nearby systems, which will be particularly important for understanding the dynamics of the lowest mass haloes in which the effects of heating are most dramatic (e.g. Fig. 5), and may also be able to measure directly the presence of outflowing gas close to the virial radius being driven out of the group potential. The sensitivity and high spatial resolution of WFI will enable high-quality temperature mapping for samples of ~100 objects (Fig. 6) so that thermodynamic conditions at the shock front can be used routinely to determine shock speeds and source age. Greatly improved X-ray inverse-Compton measurements of radio-lobe electron and magnetic field (e.g. Croston et al. 2005, Hardcastle & Croston 2010) possible with *Athena+* will provide complementary information on source dynamics and energetics for large samples.

By determining the dynamical evolution of radio galaxies as a function of source age and environment, *Athena+* will identify the locations of shock energy dissipation in the intragroup and intracluster medium for the first time. The improved understanding of radio-galaxy evolution will be crucial for the interpretation of next-generation radio surveys (e.g. by extending radio-luminosity – jet power relations to powerful sources, and by enabling the scatter introduced by environmental differences to be fully characterized) and for the realistic modelling of AGN feedback in cosmological models.

As discussed in Ettori, S., Pratt, G.W. et al. (2013, *Athena+* supporting paper) *Athena+* will map gas entropy distributions of systems over the full mass range out to $z\sim1$, identifying how the entropy excess seen in nearby clusters and groups (e.g. Pratt et al. 2010) is built up over time. The demographics of hot-gas content in low mass systems at $z>2$ will also provide stringent tests for AGN feedback models (e.g. Pointecouteau, E. Reiprich, T.H., et al., 2013, *Athena+* supporting paper). Combining these results with the major advances in radio-galaxy evolution described here will establish firmly whether AGN feedback is responsible for the cluster entropy "excess", resolving one of the major current uncertainties in cluster physics.

## 5. SUMMARY

A robust physical understanding of how AGN feedback via jets operates in galaxies, groups and clusters is an essential requirement for realistic models of galaxy evolution. This requires a major breakthrough in X-ray sensitivity and spectral resolution at high spatial resolution in order to characterize fully the thermodynamics of the hot gas phase as it interacts with and is regulated by the expanding radio lobes powered by the central AGN. Over the coming decade, a wide range of ongoing and future galaxy and cluster surveys (e.g. *Planck S-Z, Euclid, eROSITA*), and particularly radio surveys with next-generation facilities (e.g. LOFAR, SKA), will revolutionize our knowledge of galaxy and AGN populations to high redshifts. At the same time, numerical models of radio-galaxy evolution and impact are becoming increasingly sophisticated, and cosmological simulations are becoming capable of fully incorporating the physics of AGN feedback. Jet evolution models, and feedback implementations in cosmological simulations, can only be tested directly via X-ray observations of the hot gas phase on which the radio jets act. A powerful X-ray observatory such as *Athena+* has the unique ability to answer the most important outstanding questions as to how this mode of feedback operates, and should provide major breakthroughs in understanding the co-evolution of AGN, galaxies and large-scale structure.

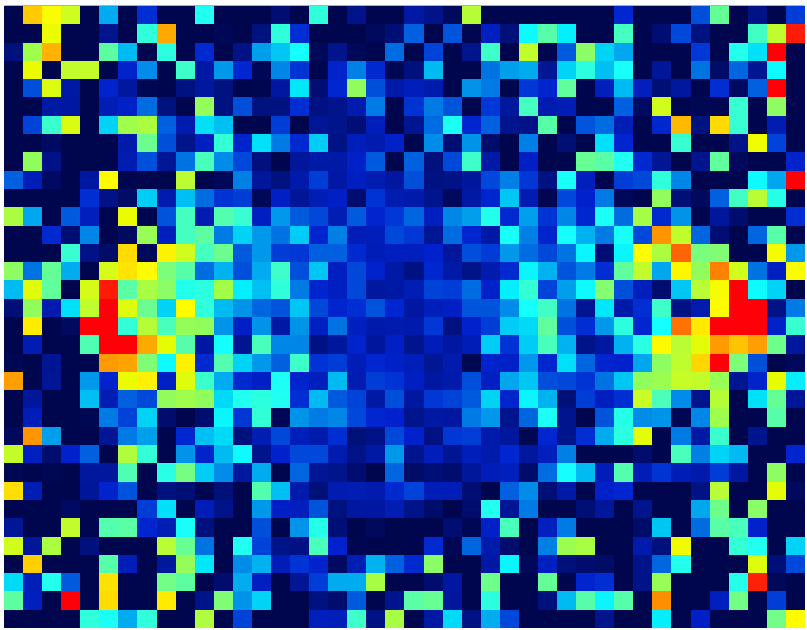

Fig. 6. Simulated WFI temperature map for a powerful radio galaxy in a group environment ($kT \sim 2$ keV) at $z=0.1$ (20-ks exposure), illustrating the characteristic temperature structures associated with the expanding bow shocks surrounding the radio lobes. The temperature ranges from ~2 keV (dark blue) to ~ 4.5 keV (red), and can be constrained to <25% in each 6"×6" region, enabling detailed tests of shock properties and source evolution for samples of 50-100 objects at intermediate redshifts. The numerical models of Hardcastle & Krause (2013) were used to generate the map, which shows a region of 135"x116".

# The Hot and Energetic Universe: AGN Feedback in Galaxy Clusters and Groups